\documentclass[
superscriptaddress,
groupedaddress,
showpacs,
twocolumn,
amsmath,
amssymb,
aps,
prb,
longbibliography,
10pt
]{revtex4-2}

\usepackage[utf8]{inputenc}
\usepackage[T1]{fontenc}
\usepackage{amsfonts}
\usepackage{graphicx}
\usepackage{soul}
\usepackage[inkscapeformat=pdf]{svg}
\usepackage[colorlinks=true,citecolor=blue,linkcolor=blue,urlcolor=blue]{hyperref}
\usepackage{float}
\usepackage{siunitx}
\usepackage{array}
\newcolumntype{C}[1]{>{\centering\arraybackslash}p{#1}}
\usepackage{newunicodechar}
\newunicodechar{∼}{\textasciitilde}
\usepackage{xcolor}

\begin{document}

\title{ 
Observation of Purcell Effect in Electrically Coupled Cavity-Magnet System
}

\author{Italo L. Soares Andrade}
\altaffiliation[Present address: ]{School of Physics and Astronomy, Cardiff University, UK}

\author{Kleber Pirota}
\affiliation{Institute of Physics Gleb Wataghin, University of Campinas (UNICAMP), 13083-859 Campinas, SP, Brazil}

\author{Amir O. Caldeira}
\affiliation{Institute of Physics Gleb Wataghin, University of Campinas (UNICAMP), 13083-859 Campinas, SP, Brazil}

\author{Francisco Rouxinol}
\email{rouxinol@unicamp.br}
\affiliation{Institute of Physics Gleb Wataghin, University of Campinas (UNICAMP), 13083-859 Campinas, SP, Brazil}

\date{\today}

\begin{abstract}
We report the observation of the Purcell effect in a cavity–metallic magnet hybrid system using electric-field–mediated coupling.
In this configuration, microwave-induced axial currents in the microwire induce circular magnetic fields that drive the ferromagnetic resonance (FMR) of the magnetized microwire. Field-dependent transmission and reflection spectroscopies reveal a clear cavity perturbation consistent with the Purcell regime, in which the magnetic loss rate exceeds the light–matter coupling strength. Despite the small magnetic volume (\(\sim 10^{-13}\,\text{m}^3\)), measurements performed at both room temperature and T=\SI{7}{\milli\kelvin} show coupling rates as high as \(g/2\pi=\SI{56}{\mega\hertz}\), one order of magnitude stronger than the one expected from conventional coupling at the magnetic antinode. Time-domain ringdown measurements directly show the magnetic-field-dependent modification of the cavity photon lifetime, in agreement with theoretical predictions. These results establish a versatile approach for coupling microwave fields to metallic magnets via geometric and electric-field-mediated interactions, opening new opportunities for hybrid cavity–magnet systems.
\end{abstract}

\maketitle
\section{Introduction}

Hybrid systems integrating microwave cavities and magnetic materials have emerged as promising platforms for applications in quantum information processing, signal transduction, detection, and memory storage~\cite{Zare_Rameshti_2022,Yuan_2022,Lachance-Quirion_Resolving_Quanta,DarkMagnonsMemory,Wendin:2017ymy}. Their appeal lies in the ability to engineer coherent interactions of microwave photons with quanta of collective spin excitations (magnons)~\cite{Tabuchi,Lachance-Quirion_Resolving_Quanta,Lachance}. In recent years, magnons have been successfully integrated into quantum device architectures based on superconducting circuitry, enabling the development of quantum magnonic platforms~\cite{Tabuchi:2015gcc,Lachance-Quirion_Resolving_Quanta,Wendin:2017ymy,Hisatomi:2016cqk,Wolski:2020mrj}.
These efforts have allowed the study of radiation-pressure-like effects~\cite{viola_kusminskiy_coupled_2016} and coupling to phononic degrees of freedom~\cite{Zhang:2016abz,Lachance-Quirion_Resolving_Quanta}, while also advancing applications in quantum transduction~\cite{Hisatomi:2016cqk}, high-sensitivity detection and quantum information processing~\cite{Wolski:2020mrj}. 
Along with developments in quantum electromechanics~\cite{Cleland:2023dsd,rouxinol_measurements_2016} and circuit quantum electrodynamics (cQED)~\cite{indrajeet_coupling_2020,wang_mode_2019,brito_testing_2015}, these cavity-magnon systems are opening new avenues for exploring quantum physics in macroscopic regimes~\cite{Tabuchi,Tabuchi:2015gcc,Zhang:2016abz}.

Most of the progress in cavity-magnonics has focused in the strong-coupling regime, in which magnons can coherently interact with microwave photons ~\cite{Huebl:2013etd,Tabuchi,Goryachev:2014ris}.
In these hybrid systems, most of the experimental efforts have largely relied on low-dissipation and high spin density materials, such as insulating ferrimagnetic yttrium iron garnet (YIG)~\cite{Tabuchi,Tabuchi:2015gcc,lachance-quirion_hybrid_2019,Lachance-Quirion_Resolving_Quanta,Zhang:2016abz}, lithium ferrite~\cite{goryachev_cavity_2018} and \(\text{Cu}_2\text{O}\text{SeO}_{3}\)~\cite{abdurakhimov_magnon-photon_2019}.
Furthermore, researchers in the field have started to explore alternative materials to expand the development of fabrication techniques that are more compatible with on-chip integration and microfabrication~\cite{li_strong_2019,hou_strong_2019}.
In particular, in these works, permalloy \(\text{Ni}\text{Fe}\) thin films have shown strong magnon-photon coupling, despite their magnetic losses~\cite{Hou}, suggesting a broader and more versatile material platform for cavity magnonics~\cite{ZHANG2023100044}.  
Recent studies have also explored alternative coupling regimes that have stimulated new ideas to engineer and investigate applications thereof, such as magnetically induced transparency \cite{liu_optomagnonics_2016} and the Purcell effect~\cite{Zhang_2014,MIT-YIG-PY-bilayers,VermaPurcell}. The latter, in particular, underpins a variety of applications in cavity electrodynamics, ranging from single-photon sources~\cite{single_mw_photonsource,purcell_qdot_singlephoton}, and the increased laser efficiency~\cite{8281504}, to qubit relaxation protection~\cite{Suppresionqubitrelaxationpurcell}, also shining  light on the dissipative dynamics in such regimes~\cite{zhao2025theorymagnonpurcelleffect,PhysRevResearch.6.023052}. These results show the importance of understanding and controlling both the coupling constant and dissipative properties of metallic magnetic materials, particularly those with significant losses such as metallic ferromagnets.

However, despite recent progress in the development of hybrid systems incorporating magnetic materials, metallic magnetic systems remain comparatively underexplored, although they play a key role in enabling a broad range of physical phenomena, from the fundamental investigation of quantum effects to technological applications of magneto-impedance~\cite{CHIRIAC1996333,Pirota}, spin current manipulation~\cite{bai_cavity_2017}, and long-range cavity-mediated magnon coupling~\cite{janssonn_cavity-mediated_2023}.

In contrast to insulating magnets, metallic microwires can exhibit ferromagnetic resonance (FMR) driven by the cavity’s electric field, a mechanism first demonstrated by Rodbell in iron whiskers~\cite{Rodbell_Whisker}. When placed at the electric field antinode, such wires behave as dipole antennas that concentrate the microwave field in their immediate surroundings. The resulting current, confined within the skin depth of the conductor, generates a localized magnetic field that can greatly exceed the native magnetic field of the cavity~\cite{Rodbell_ConductingPertubation}, giving high FMR signals.

Here, we report measurements of a cavity-magnet system composed of a metallic glass-coated amorphous CoFeSiB microwire~\cite{Pirota,CHIRIAC1996333} and a 3D microwave cavity operated at both ambient and cryogenic temperatures. We detect a coupling between a magnetic mode in a microwire and an electromagnetic field of a cavity mediated by its electric field, in an antenna-like configuration. We perform frequency-domain spectroscopy varying the magnetic field bias around the ferromagnetic resonance and extract the coupling strength and dissipation rates, confirming that the system is in the Purcell regime. We further explore the regime through time-resolved cavity ringdown and extract the magnetic field-dependent photon lifetime at cryogenic conditions. Our results indicate that even high-loss magnetic materials can be effectively studied via cavity magnonic models and that the electrically mediated coupling can be incorporated in future quantum and classical hybrid microwave architectures.

\section{Theory}
\subsection{Cavity-Magnet coupling strength and the Purcell Regime}

In the linear regime, cavity–magnet hybrid systems are commonly described as two coupled harmonic oscillators, one representing a single-mode electromagnetic field of a cavity and the other, the ferromagnetic precession mode. Within the rotating-wave approximation, the system Hamiltonian takes the form~\cite{CavityMagnonics}:
\begin{equation}
    \mathcal{H} = \hbar \omega_c a^\dagger a + \hbar \omega_m m^\dagger m + \hbar g (a^\dagger m + a m^\dagger),
    \label{eq:Hamiltonian}
\end{equation}
where \( \omega_c \) and \( \omega_m \) respectively denote the resonance frequencies of the bare cavity and the magnon mode, \( a^{\dagger} \) (\(a\)) and \( m^{\dagger} \) (\(m\)) are boson operators that create (annihilate) the photon and magnon modes, respectively, and \( g \) is the coupling strength. 
In conventional cavity magnonics, the coupling is governed by the overlap between the magnetic mode (magnon) and the zero-point fluctuations of the magnetic field, \(\mathbf{B}_{ZPF}\), and scales with the square root of the number of spins \(\sqrt{N}\)~\cite{Lachance}. For a uniformly magnetized sample under uniform field, we have:
\begin{equation}
g=\gamma B_{ZPF}\sqrt{N},
\end{equation}
where \(\gamma\) is the gyromagnetic ratio. This scaling motivates the use of large-volume magnetic samples and high-field-density cavity geometries to enhance the coupling rate.

\begin{figure}[ht]
    \centering
    \includegraphics[width=1\linewidth]{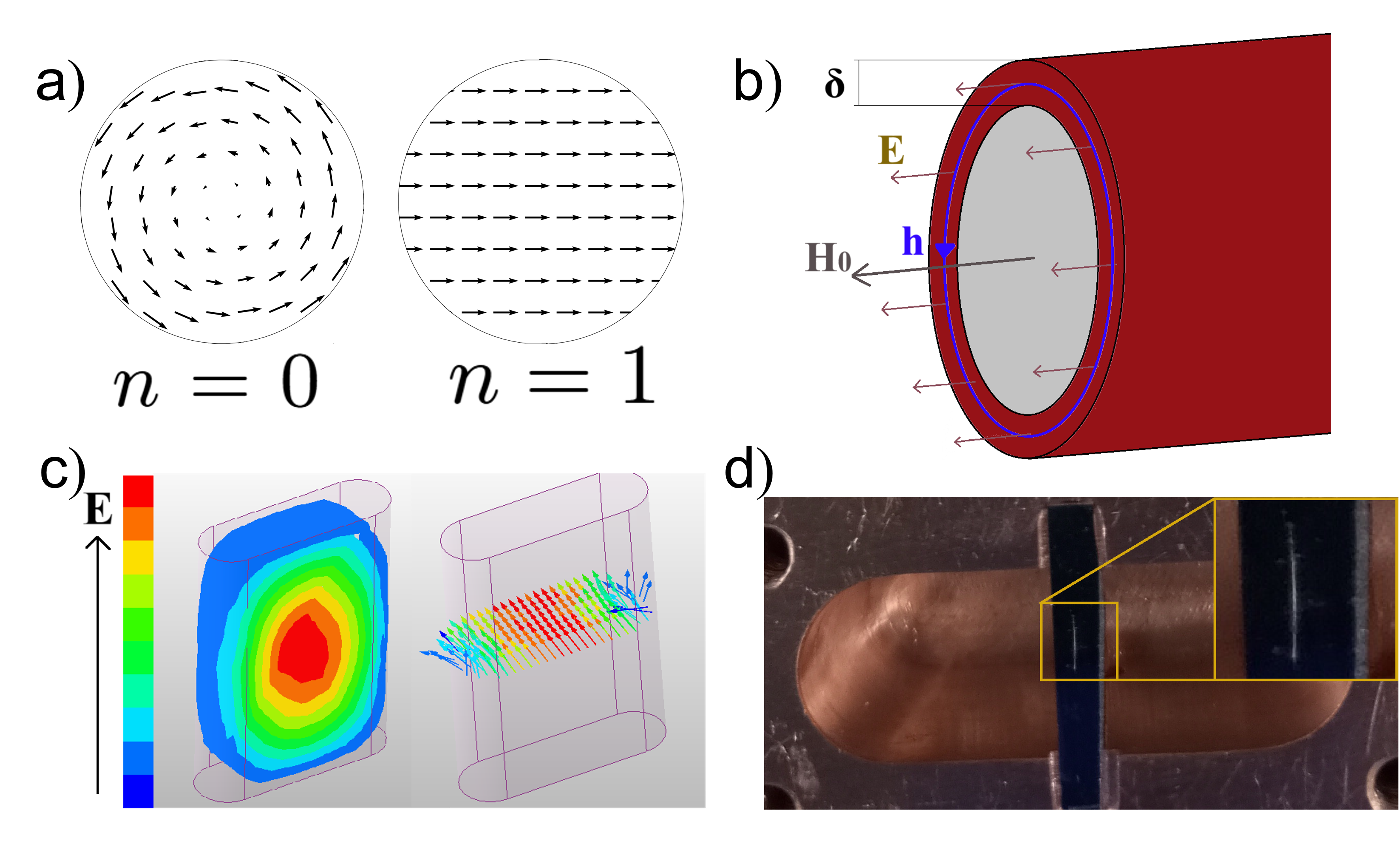}
    \caption{ 
    (a) Illustration of the ac magnetization in the circular mode \(n=0\) and in the dipolar mode \(n=1\).
    (b) An oscillating current, represented in red colour, generates a circular field $\mathbf{h}$ that couples with mode \(n=0\). 
    (c) Ansys HFSS~\cite{ansysHFSS} simulation of the electric field distribution of the electromagnetic TE$_{101}$ with the mode in a rectangular cavity (\SI{26}{\milli\meter} × \SI{8}{\milli\meter} × \SI{36}{\milli\meter}), showing the electric antinode in the middle. 
    d) A \SI{2.5}{\milli\meter} piece of wire affixed on silicon mounted on the electric antinode of the cooper cavity.
    }
    \label{fig:modes}
\end{figure}

The response of the coupled cavity–magnet system can be described using the input–output formalism~\cite{Milburn}, which relates the transmitted and reflected signals to the internal dynamics of the system. In terms of the probe frequency \(\omega\), the complex transmission coefficient \( S_{21}(\omega) \) is given by~\cite{Tabuchi}:
\begin{equation}
    S_{21}(\omega) = \frac{\sqrt{\kappa_{1,\mathrm{ex}} \kappa_{2,\mathrm{ex}}}}{i(\omega - \omega_c) - \kappa_c/2 + \dfrac{g^2}{i(\omega - \omega_m) - \kappa_m/2}},
    \label{eq:S21}
\end{equation}
where \( \kappa_c \) is its total cavity decay rate (including internal and external losses) and \( \kappa_m \) is the dissipation rate of the magnetic system, both related to the linewidth of the respective system. The parameters \( \kappa_{1,\mathrm{ex}} \) and \( \kappa_{2,\mathrm{ex}} \) represent the external coupling rates at the input and output ports, respectively. The reflection coefficient for the output port is given by~\cite{Pozar}: 
\begin{equation}
    S_{22}=1+\sqrt{\frac{\kappa_{2,ex}}{\kappa_{1,ex}}}S_{21}.    
\end{equation}

When both dissipation rates \( \kappa_c\), \( \kappa_m\) are smaller than the coupling rate \(g\), the system is in the strong-coupling regime; Rabi splitting occurs at the resonance and energy coherently oscillates between the cavity and the magnet \cite{Zhang_2014,zhao2025theorymagnonpurcelleffect}. On the other hand, when
\begin{equation}
    \kappa_m \gg g \gg \kappa_c \label{eq:RegimePurcell},
\end{equation}
there is no Rabi splitting and the cavity spectrum remains single-peaked. The resonance frequency and linewidth of the cavity are perturbed according to~\cite{Abe}:
\begin{align}
    \omega_{\mathrm{sys}} &= \omega_c - \frac{g^2 \Delta}{\Delta^2 + (\kappa_m/2)^2}, \label{eq:shift} \\
    \kappa_{\mathrm{sys}} &= \kappa_c + \frac{g^2 \kappa_m}{\Delta^2 + (\kappa_m/2)^2},
    \label{eq:broadening}
\end{align}
where \( \Delta = \omega_m - \omega_c \) is the detuning between the magnetic and photonic modes. 

In the regime given by Equation~\ref{eq:RegimePurcell}, the system can be perturbatively treated; Equation~\eqref{eq:broadening} being an expression of the Fermi Golden rule, where the cavity decay rate is increased by the additional density of states provided by the lossy magnetic system (see Appendix~\ref{appendix:purcell_regime}). Hence, Equation~\ref{eq:RegimePurcell} determines the so called Purcell regime~\cite{Zhang_2014}: the cavity, that has the role of the "emitter", suffers the Purcell effect caused by its interaction with the lossy magnetic system. By contrast, when the roles of $\kappa_c$ and $\kappa_m$ are inverted, \textit{i.e.}, and the cavity is more lossy than the magnet, the system enters the magnetically-induced transparency regime, and the Purcell effect happens in the magnet instead. Equations~\eqref{eq:shift} and~\eqref{eq:broadening} can be derived from the coupled harmonic oscillators model~\eqref{eq:Hamiltonian} in the limit of high dissipation; we refer to Appendix~\ref{appendix:purcell_regime} for details. 

The time-domain dynamics of the coupled cavity–magnet system described by Eq.~\ref{eq:Hamiltonian} can be modelled using the Lindblad master equation, with dissipation incorporated via the jump operators \( \sqrt{\kappa_c}a \) and \( \sqrt{\kappa_m}m \). In the undriven case, the expectation value of the operators follows~\cite{zhao2025theorymagnonpurcelleffect}:
\begin{equation}
    \frac{d}{dt}
    \begin{pmatrix}
        \langle a^\dagger a \rangle \\
        \langle m^\dagger m \rangle \\
        \langle a^\dagger m \rangle \\
        \langle a m^\dagger \rangle
    \end{pmatrix} = \hat{\Omega}
    \begin{pmatrix}
        \langle a^\dagger a \rangle \\
        \langle m^\dagger m \rangle \\
        \langle a^\dagger m \rangle \\
        \langle a m^\dagger \rangle
    \end{pmatrix},
    \label{eq:evolution}
\end{equation}
where $\hat{\Omega}$ is given by: 
\begin{equation}
    \hat{\Omega} = 
    \begin{pmatrix}
        -\kappa_c & 0 & -ig & ig \\
        0 & -\kappa_m & ig & -ig \\
        -ig & ig & -i\Delta - \frac{\kappa_c + \kappa_m}{2} & 0 \\
        ig & -ig & 0 & i\Delta - \frac{\kappa_c + \kappa_m}{2} 
    \end{pmatrix}.
    \label{eq:LindbladC}
\end{equation}
This expression gives a combination of oscillatory and decaying terms. In the Purcell regime (\( \kappa_m \gg g \gg \kappa_c \)), the decay terms dominate the dynamics. When one real exponential component dominates the decay of the intra-cavity photon number, \(\langle a^\dagger a \rangle \propto e^{-t/\tau}\), it is possible to define the photon lifetime \(\tau\).  

\subsection{Ferromagnetic Resonance in Thin Metallic Wires}\label{sec:FMR-thin_metalic}

Ferromagnetic resonance in metallic wires reflects the interplay between the wire's geometry and its finite conductivity, leading to cylindrical magnetic modes~\cite{Kraus2011} and additional damping due to eddy currents and associated skin effect.
The system's dynamics is jointly described by Maxwell and Landau–Lifshitz equations~\cite{AmentRado}, which determine the dispersion relation \( k(\omega) \) and the effective magnetic permeability tensor~\cite{KRAUS2015449}.
In the situation where the wire is magnetised along its axis and the microwave magnetic fields are transverse, the time-varying component of the magnetisation in the cylindrical geometry is given by~\cite{Kraus2011}:
\begin{equation}
    \mathbf{m}(r, \phi, t) = e^{i\omega t} \sum_n \mathbf{m}_n\, e^{i n \phi} J_n(k r), 
    \label{cylindrical_modes}
\end{equation}
where \(n\) denotes the azimuthal mode number, \( J_n(x) \) is the Bessel function of the first kind, and \(\omega\) refers to the frequency of the driving electromagnetic field. 

The FMR response depends on the symmetry of the incident electromagnetic field, which, together with the wire's boundary conditions, determines the set of excited cylindrical modes in the system~\cite{Kraus1982}. Among these, the modes \( n = 0\) and \( n = 1 \) are particularly relevant. The \( n = 0 \) mode exhibits a circular symmetry [see Fig.~\ref{fig:modes}(a)], and is predominantly excited by a circular magnetic field, such as those generated by radio-frequency (RF) electric currents flowing along the wire axis. In contrast, the \( n = 1\) mode tends to have a dipole character, particularly when the skin depth is larger than the wire radius, resembling the FMR mode of insulating cylinders.
From the circular symmetry of the \( n = 0 \) mode, its resonance frequency is given by the Kittel resonance condition for a uniform magnetised plane~\cite{kittel1948}:
\begin{equation}
    \omega_m = \gamma \sqrt{B_0(B_0 + \mu_0 M_s)},
    \label{eq:Kittel}
\end{equation}
where \( B_0\) is the applied static field and \( M_s\) is the saturation magnetisation. 

When the wire is exposed to electric fields, microwave currents are induced along its length, constrained by the skin-depth near the surface as illustrated in Fig.~\ref{fig:modes}(b). These axial currents generate strong circular magnetic fields around the wire, as dictated by Ampère's law, with magnetic field amplitudes scaling as \( h \sim I/2\pi R\), where \(R\) is the wire's radius. As a consequence, the mode \( n = 0 \) is usually the dominant excitation in FMR of metallic microwires~\cite{Kraus2011}. This mechanism was first recognized by Rodbell in the 1950s~\cite{Rodbell_Whisker,Rodbell_ConductingPertubation}, who demonstrated that placing the wire at the electric field antinode, rather than at the conventional magnetic field antinode, significantly enhance the FMR signal in cavity measurements. In this configuration, the magnetic fields generated near the wire surface can exceed the cavity magnetic field by many orders of magnitude. 

It is indeed this electric-field-mediated coupling that we investigate in the present work. By positioning the microwire at the electric field antinode of a 3D microwave cavity [Figs.~\ref{fig:modes}(c) and (d)], we probe the coupling of the \( n=0 \) magnetic mode to the cavity's electromagnetic field in the Purcell regime; the system or experimental data is analysed within the framework of the coupled harmonic oscillator model [Eq. \ref{eq:Hamiltonian}].

\section{Experimental Methods}
\subsection{Magnetic microwires}

\begin{figure}[h]
    \centering
    \includegraphics[width=1\linewidth]{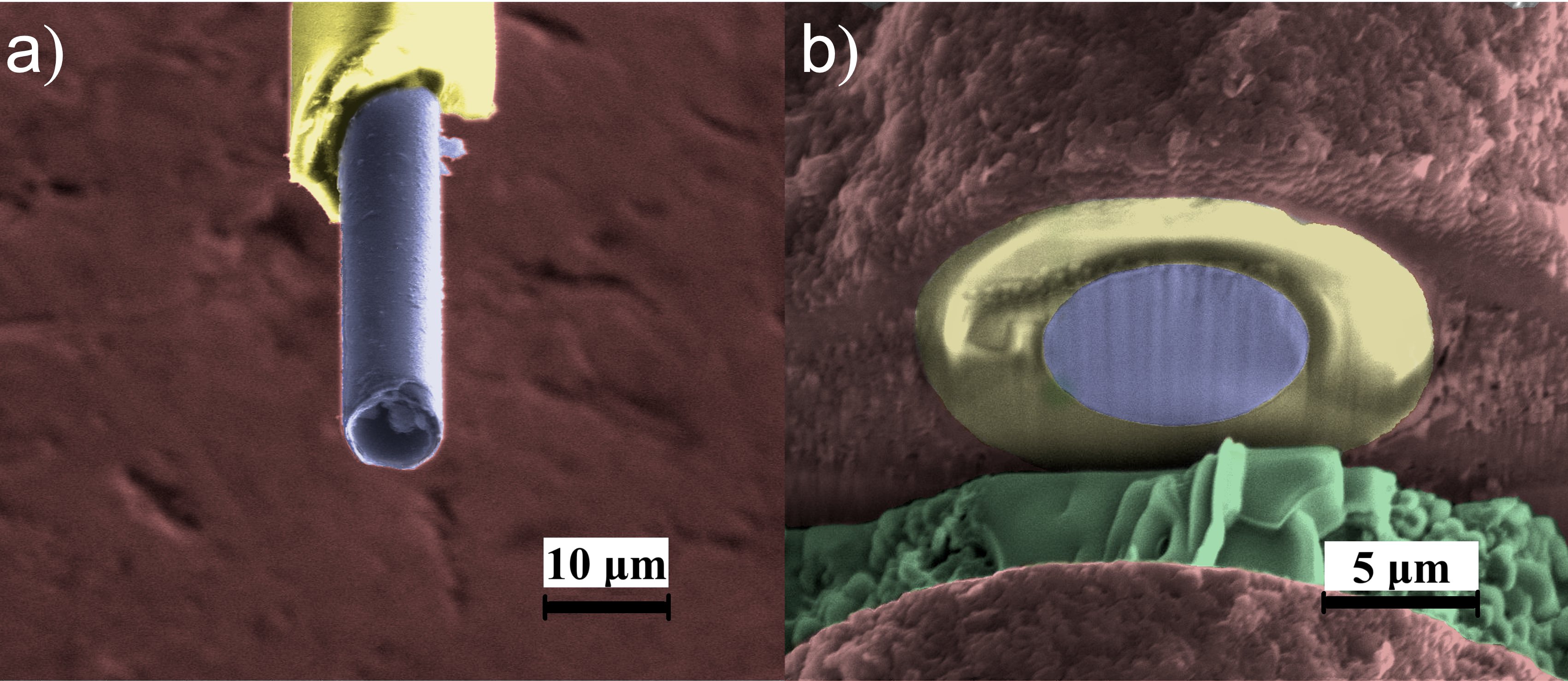}
    \caption{ 
    False-colour SEM images of CoFeSiB microwires. 
    (a) Image of the wire tip, showing the exposed metallic core and the surrounding glass cladding. 
    (b) Cross-sectional view of a wire embedded in carbon and sectioned by FIB milling. The metallic core, with a diameter of \SI{8.5}{\micro\metre}, is clearly visible and surrounded by the insulating glass layer. Tilt angle: 30\(^\circ\).
    }
    \label{fig:SEM-FIB}
\end{figure}

The magnetic samples used in this study are amorphous CoFeSiB microwires fabricated using the Taylor–Ulitovsky method~\cite{chemosensors10010026}, which yields continuous flexible wires coated with a glass insulating layer. This cladding enhances mechanical robustness and provides electrical isolation. The nominal composition is \(\mathrm{Co}_{68.15}\mathrm{Fe}_{4.35}\mathrm{Si}_{12.5}\mathrm{B}_{15}\), with a metallic core radius of approximately \SI{4}{\micro\meter} and a total outer diameter of \SI{16}{\micro\meter}, see Fig.~\ref{fig:SEM-FIB}(a). Figure~\ref{fig:SEM-FIB}(b) shows a false-coloured scanning electron microscopy (SEM) image of a representative cross-section, prepared by focused ion beam (FIB) milling, distinguishing the high-contrast interface between the metallic core and the surrounding glass. For imaging, the wire was mounted on an aluminium stub and electrically grounded using carbon paint to mitigate charging under electron irradiation.

The magnetic properties of the microwires were characterized at room temperature using a vibrating sample magnetometer (VSM). The samples exhibit soft magnetic behaviour, with coercive fields below \SI{10}{\text{Oe}} and saturation magnetisation values in the range of \(\mu_0 M_s = 0.81\)-\SI{0.85}{\tesla}. The electrical resistivity is of the order \(\sim 10^{-6}\) \SI{}{\ohm\meter} \cite{CHIRIAC1996333}.

\subsection{Sample preparation}

Microwires with lengths ranging from \SI{2.5}{\milli\meter} to \SI{5}{\milli\meter} were selected and mounted on dielectric substrates tailored for either room-temperature (RT) or cryogenic measurements. At room temperature, paper substrates were employed for simplicity, with wires affixed using double-sided adhesive tape. For cryogenic measurements, the wire was placed on high-resistivity silicon die (\(> \SI{10}{\kilo\ohm\centi\meter}\)) using cryogenic varnish \footnote{(GE/IMI-7031 Varnish)} to ensure mechanical stability and improve thermal anchoring. 

In both configurations, the wires were positioned such that their axis was parallel to both the cavity electric field and the externally applied static magnetic field \(\mathbf{B}_{0}\) in order to maximize the coupling strength to the cavity electric field and suppress excitations to unwanted modes, see Fig.~\ref{fig:modes} (c) and (d).

\subsection{Cavity Characteristics}
\begin{figure}[h]
    \centering
    \includegraphics[width=1\linewidth]{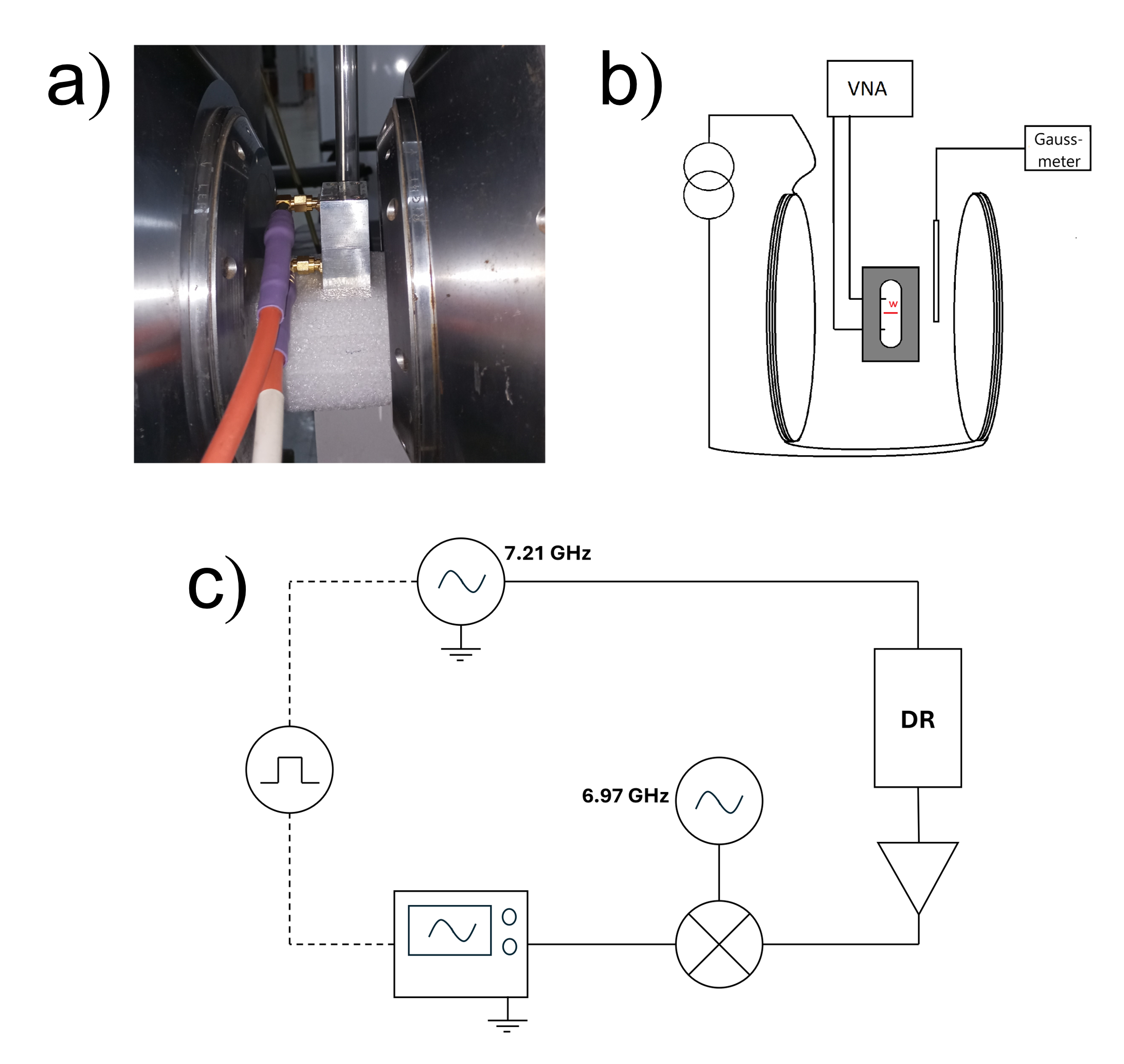}
    \caption{
    (a) Image of the room-temperature measurement setup. The cavity is connected to the vector network analyser (VNA) via coaxial cables and supported by a foam platform placed between the poles of an electromagnet. A Hall probe behind the sample is used for field calibration. 
    (b) Schematic diagram of the room-temperature microwave setup showing the current source, the cable to the VNA, and the Gaussmeter. 
    (c) Schematic of the cryogenic cavity ringdown setup. The AWG modulates an RF source to generate microwave pulses that are delivered to the sample. The transmitted signal is down-converted to \SI{240}{\mega\hertz} and recorded by a fast oscilloscope for time-domain analysis.
    }
    \label{fig:sup-RT_and_FAST_Aquisition}
\end{figure}

The microwave cavities used in this study were rectangular 3D resonators machined from either 6061 aluminium alloy or oxygen-free high-conductivity (OFHC) copper. All cavities were designed to operate in the \(\mathrm{TE}_{101}\) mode, with bare resonance frequencies ranging from 7.2 to \SI{7.4}{GHz}. 

Microwave coupling was implemented through SMA connectors weakly coupled to the cavity through cylindrical antenna pins. This configuration enables transmission (\(S_{21}\)) measurements at both RT and cryogenic temperatures, as well as reflection (\(S_{22}\)) measurements under RT conditions.

\begin{figure*}[ht]
    \centering
\includegraphics[width=\textwidth,keepaspectratio]{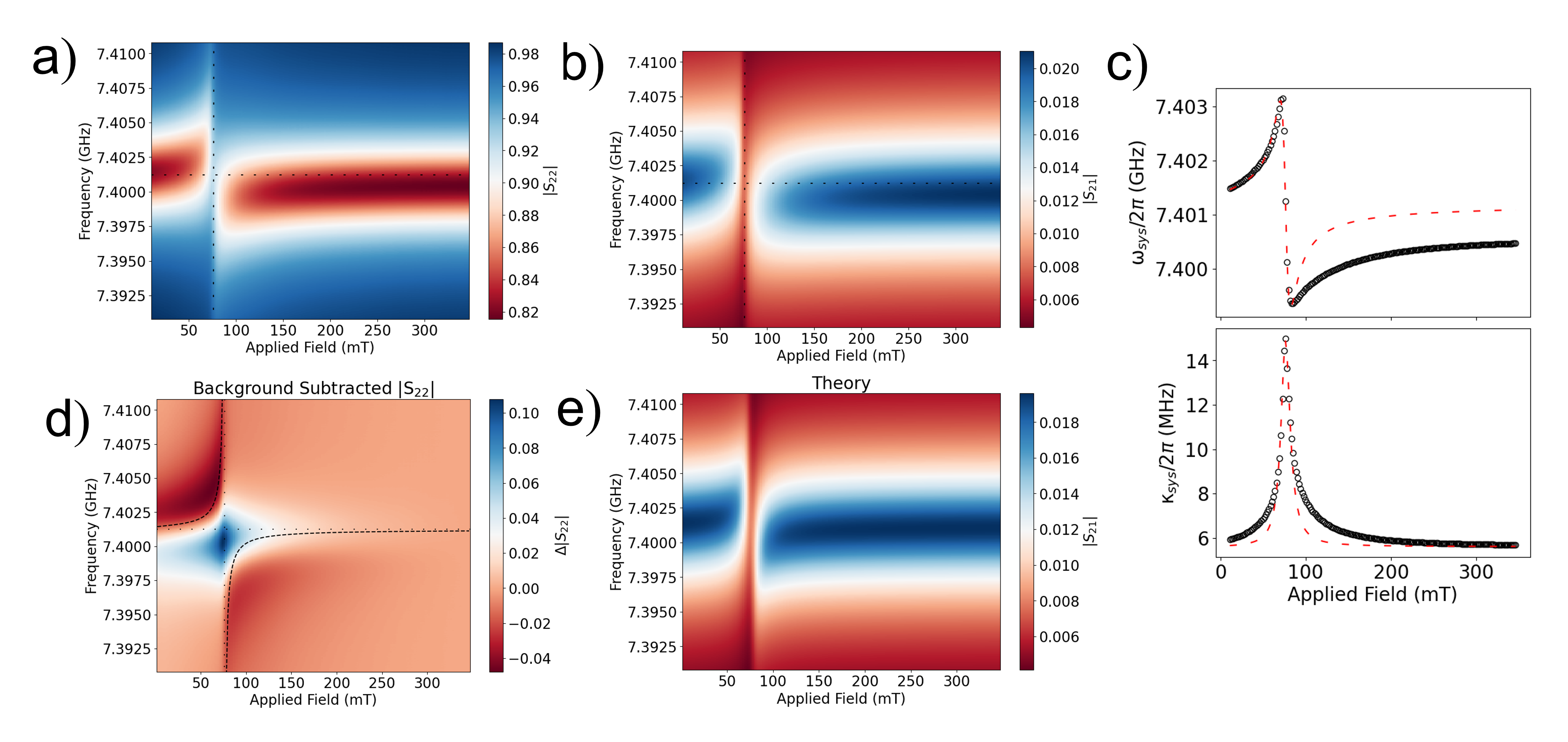}
    \caption{
    Magnetic field-dependent response of the cavity coupled to a \SI{4}{\milli\meter} CoFeSiB microwire positioned at the electric field antinode. 
    (a) Reflection spectrum (\(S_{22}\)) as a function of static magnetic field, showing linewidth broadening near \SI{76}{mT}, consistent with the onset of FMR. 
    (b) Transmission spectrum (\(S_{21}\)), under identical conditions as in (a), exhibiting absorption near resonance with reduced contrast due to asymmetric port coupling. 
    (c) Extracted resonance frequency and linewidth as a function of magnetic field. The red dashed line corresponds to the model prediction from Equations~\eqref{eq:shift} and ~\eqref{eq:broadening} using fitted values of \(g\) and \(\kappa_m\). 
    (d) Background subtracted reflection spectra (\(|S_{22}(B_0)|-|S_{22}(\SI{345}{\milli\tesla})|\))  showing the changes caused by the coupling with the magnet. The polariton eigenfrequecies of Equation~~\eqref{eq:Hamiltonian} are plotted in dashed lines as a reference. (e) Transmission spectra calculated from Equation~\eqref{eq:S21} using the extracted parameters. }
    \label{fig:4mm}
\end{figure*}

\subsection{Measurement setup}

Room-temperature measurements were performed using a Helmholtz coil to generate a static magnetic field, calibrated via a Hall sensor positioned near the cavity [Figs.~\ref{fig:sup-RT_and_FAST_Aquisition}(a) and (b)]. Cryogenic measurements were carried out in a commercial dry dilution refrigerator with a base temperature of \SI{10}{mK}, equipped with a superconducting magnet anchored to the \SI{4}{\kelvin} stage (see Appendix~\ref{appendix:cryogenic_measurement} for details). The OFHC copper cavity was anchored to the mixing chamber (MXC) and aligned to the centre of the magnetic field. 

Spectroscopy at RT was conducted using a vector network analyzer (VNA), calibrated using a standard Short-Open-Load-Thru (SOLT) protocol~\cite{williams_optimal_2003}, with an input power of \SI{1}{\milli\watt} (\SI{0}{dBm}). For cryogenic measurements, cavity response was both recorded in frequency and time domains. The latter employed a home-made superheterodyne detection system for ringdown spectroscopy (see Appendix~\ref{Appendix:ringdown}).

Figure~\ref{fig:sup-RT_and_FAST_Aquisition}(c) shows a schematic representation of the heterodyne circuit used to probe the time-resolved photon decay. Pulsed signals (\SI{200}{\micro\second}) at the cavity frequency were applied at fixed values of \(B_{0}\); the transmitted response was down-converted to an intermediate frequency of \SI{240}{\mega\hertz} and digitized using a high-speed oscilloscope triggered by an arbitrary waveform generator. 

\section{Results}

\subsection{Frequency-Domain Spectroscopy at Room Temperature}

Figures~\ref{fig:4mm} (a) and (b) show RT measurements of the frequency-dependent transmission and reflection spectra as a function of the applied magnetic field for a \SI{4}{\milli\meter} long wire in the aluminium cavity. 
The bare resonance frequency of the cavity was \SI{7.435}{\giga\hertz}, but in the presence of the wire, it is shifted to \SI{7.401}{\giga\hertz} due to the modification of the cavity’s electromagnetic environment. A pronounced absorption is observed near \SI{76}{\milli\tesla}, consistent with the FMR of the microwire. 
From the Kittel relation~\eqref{eq:Kittel}, the measured resonance field and frequency correspond to a saturation magnetization of \(\mu_0 M_s = \SI{0.84}{\tesla}\), which agrees with the values obtained independently from VSM measurements. Additionally, the frequency response exhibits a single broadened Lorentzian lineshape, without any mode splitting, indicating that the system is in a weak coupling regime. 

Fits of the power transmission spectra \(|S_{21}|^2(\omega)\) to Lorentzian functions at each magnetic field value allow us to extract the field-dependent cavity linewidths and resonance frequencies, as shown in Figures~\ref{fig:4mm}(c).
It is evident from the data the linewidth broadening and the frequency shift near the resonance. An independent analyses (see Appendix~\ref{Appendix:Parameter_Cal}) with the aid of Eqs.~\eqref{eq:shift} and~\eqref{eq:broadening} (dashed red lines) yields coupling rates of \(g/2\pi \simeq \SI{35}{\mega\hertz}\) and \(g/2\pi \simeq \SI{40}{\mega\hertz}\) and magnetic dissipation rates of \(\kappa_m/2\pi \simeq \SI{640}{\mega\hertz}\) and \(\kappa_m/2\pi \simeq \SI{690}{\mega\hertz}\), respectively. The basis of the linewidth curve yields a non-magnetic cavity decay rate of \(\kappa_c/2\pi \sim \SI{5.6}{\mega\hertz}\). Such a difference in the values of the coupling rate \(g/2\pi\) and the magnetic dissipation \(\kappa_m/2\pi\) was observed in all measurements; we then consider the mean (or average) values as the representative ones. We plot the \(\omega_c\) and \(\omega_m\) in dotted lines.

To further understand the experimental behaviour, we isolate the coupling contribution in the reflection data by subtracting the off-resonant cavity response~\cite{Mawgan} at \SI{346}{\milli\tesla}, see Fig.~\ref{fig:4mm}(d); in this plot, we also include the eigenvalues calculated from the coupled  harmonic oscillator model~(\ref{eq:Hamiltonian}) using the mean coupling rate \(g/2\pi=\SI{37}{\mega\hertz}\) (dashed black lines).

Figure~\ref{fig:4mm}(e) shows the numerically simulated transmission spectra \(|S_{21}|\) determined via input–output theory [see Eq.~\ref{eq:S21}] with the extracted parameters and \( \kappa_1=\SI{5.8}{kHz}\) and \(\kappa_2 = \SI{540}{kHz}\). The simulated spectra reproduce the general spectral features observed in the experimental one. 
Despite the overall agreement with the coupled-mode model~(\ref{eq:Hamiltonian}), deviations in the lineshapes are evident, suggesting additional effects beyond the ideal harmonic oscillator approximation. Nevertheless, the Purcell-regime coupled oscillator model offers a compact and effective description of the electric-field-driven FMR in metallic microwires.

\subsection{Dependence on Position and Wire Length}
\begin{figure}[!h]
    \centering
    \includegraphics[width=1\linewidth]{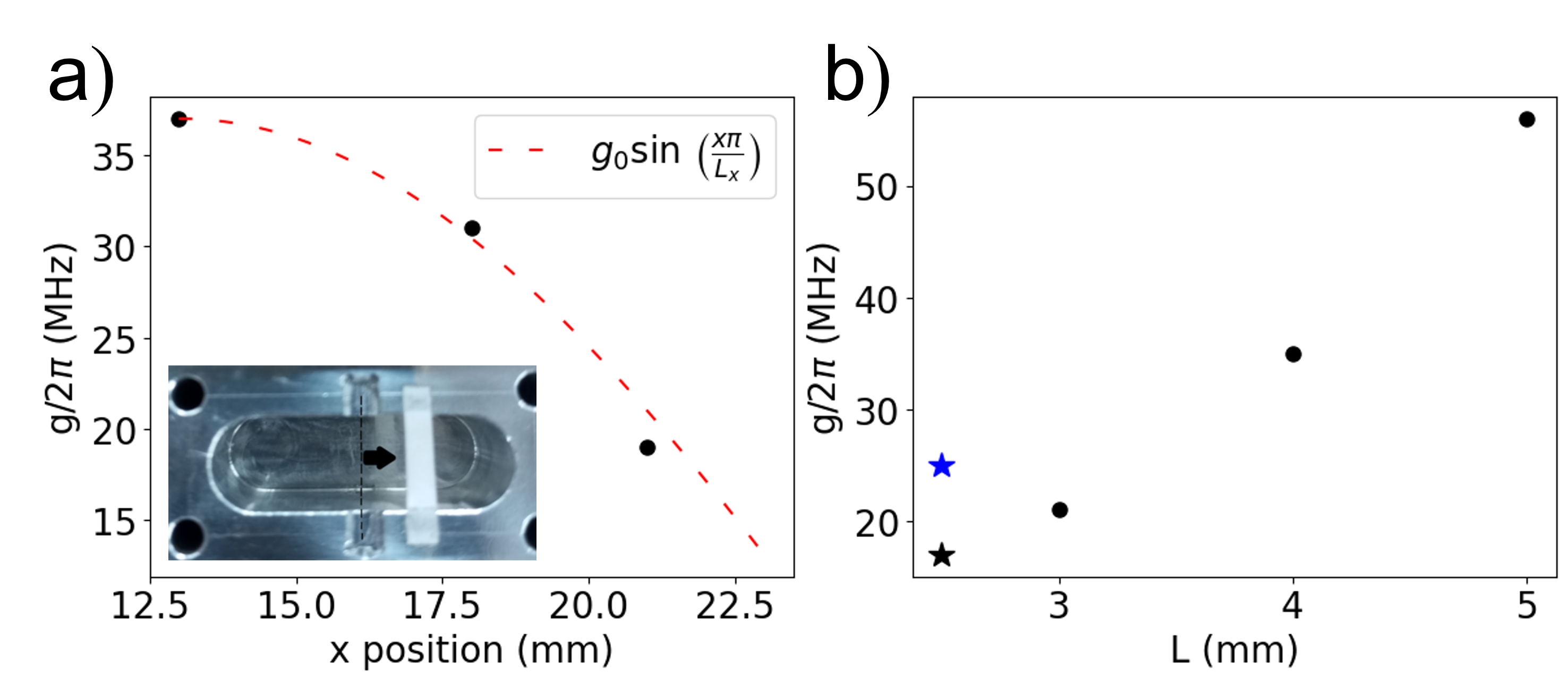}
    \caption{
    (a) Coupling strength \( g/2\pi \) as a function of the wire's lateral displacement from the centre of the cavity along the \( x \)-axis. The red dashed line represents the expected sinusoidal dependence of the electric field in the \( \mathrm{TE}_{101} \) mode, normalized to the value at the electric antinode. The inset shows a photograph of the aluminium cavity and illustrates the sample displacement direction.
    (b) Coupling strength as a function of wire length L for different substrate and cavity configurations. Black markers correspond to measurements performed in the aluminium cavity: circles represent samples mounted on paper substrates, and stars represent samples mounted on high-resistivity silicon dies. The blue star indicate a measurement performed in a copper cavity.
    }
    \label{fig:g-variation}
\end{figure}

To verify that the coupling is indeed mediated by the cavity's electric field, we measured the coupling strength as a function of the lateral displacement of the wire from the cavity's electric field antinode. As shown in Fig.~\ref{fig:g-variation}(a), the extracted coupling strength decreases monotonically as the wire is moved away from the electric antinode. Besides, the values closely follow the spatial profile of the electric field in the \(\mathrm{TE}_{101}\) mode, which varies sinusoidally along the cavity axis. A normalized sine function is plotted as a guide for the eyes (dashed red line).

Measurements were also made exploring the variation of the coupling strength with the wire length, shown in Fig.~\ref{fig:g-variation}(b). For samples with identical mounting conditions (paper substrate and aluminium cavity), the coupling exhibits a monotonic increase with the wire length (see Sec.~\ref{sec:Discussion} for details). Samples mounted on high-resistivity silicon dies and measured in different cavities show deviations, indicating that the coupling is sensitive to the electromagnetic environment. These and other measured parameters are summarized in Table \ref{tab:table_data}.

\subsection{Low-Temperature Measurements and Cavity Ringdown}

\begin{figure}[h]
    \centering
    \includegraphics[width=1\linewidth]{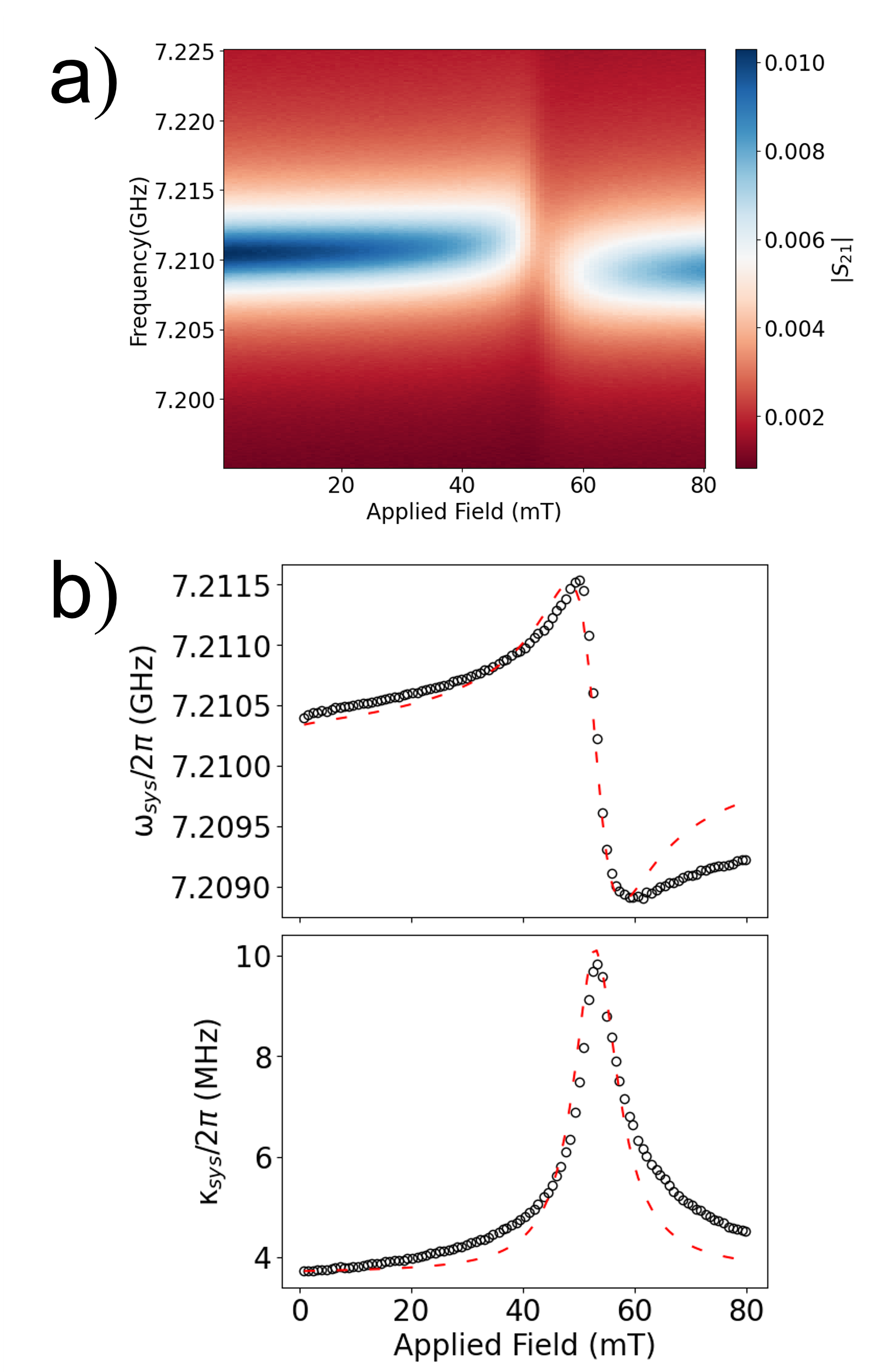}
    \caption{ 
    (a) Frequency-domain cavity transmission at \SI{7}{\milli\kelvin} for a \SI{2.5}{\milli\meter} CoFeSiB wire. 
    (b) Fitted cavity resonance and linewidth as a function of the applied field. 
    }
    \label{fig:Low-T-meas}
\end{figure}

Figure~\ref{fig:Low-T-meas}(a) shows the magnetic field-dependent transmission of the \SI{2.5}{\milli\meter} wire measured inside a copper cavity at \(T=\SI{7}{\milli\kelvin}\). A pronounced absorption feature is observed near \SI{53}{\milli\tesla}, consistent with FMR. According to the Kittel relation~\eqref{eq:Kittel}, such a feature corresponds to a saturation magnetization of \(M_s = \SI{1.19}{\tesla}\), reflecting an increase in magnetization due to reduced thermal fluctuations~\cite{swierczek_temperature_1988}. Using the same fitting procedure applied to the RT \(S_{21}\) spectra, we extract the relevant parameters \(g/2\pi\), \(\kappa_m/2\pi\) and \(\kappa_c/2\pi\), summarized in Table~\ref{tab:table_data}. 
We find that both dissipative rates exhibit  minor variation with temperature: \(\kappa_m/(2\pi) \simeq \SI{680}{\mega\hertz}\) at \(T=\SI{7}{\milli\kelvin}\) and \(\kappa_m/2\pi \simeq \SI{660}{\mega\hertz}\) at RT; \(\kappa_c/2\pi \simeq \SI{3.7}{\mega\hertz}\) at \(T=\SI{7}{\milli\kelvin}\) and \(\kappa_c/2\pi \simeq \SI{4.1}{\mega\hertz}\) at RT. These results suggest that the dominant loss mechanisms are largely temperature-independent over the range explored. The coupling strength exhibits a slightly increasing from  \(g/2\pi \simeq \SI{25}{\mega\hertz}\) at room temperature to \(g/2\pi \simeq \SI{32}{\mega\hertz}\) at cryogenic conditions.

\begin{figure}[ht]
    \centering
    \includegraphics[width=1\linewidth]{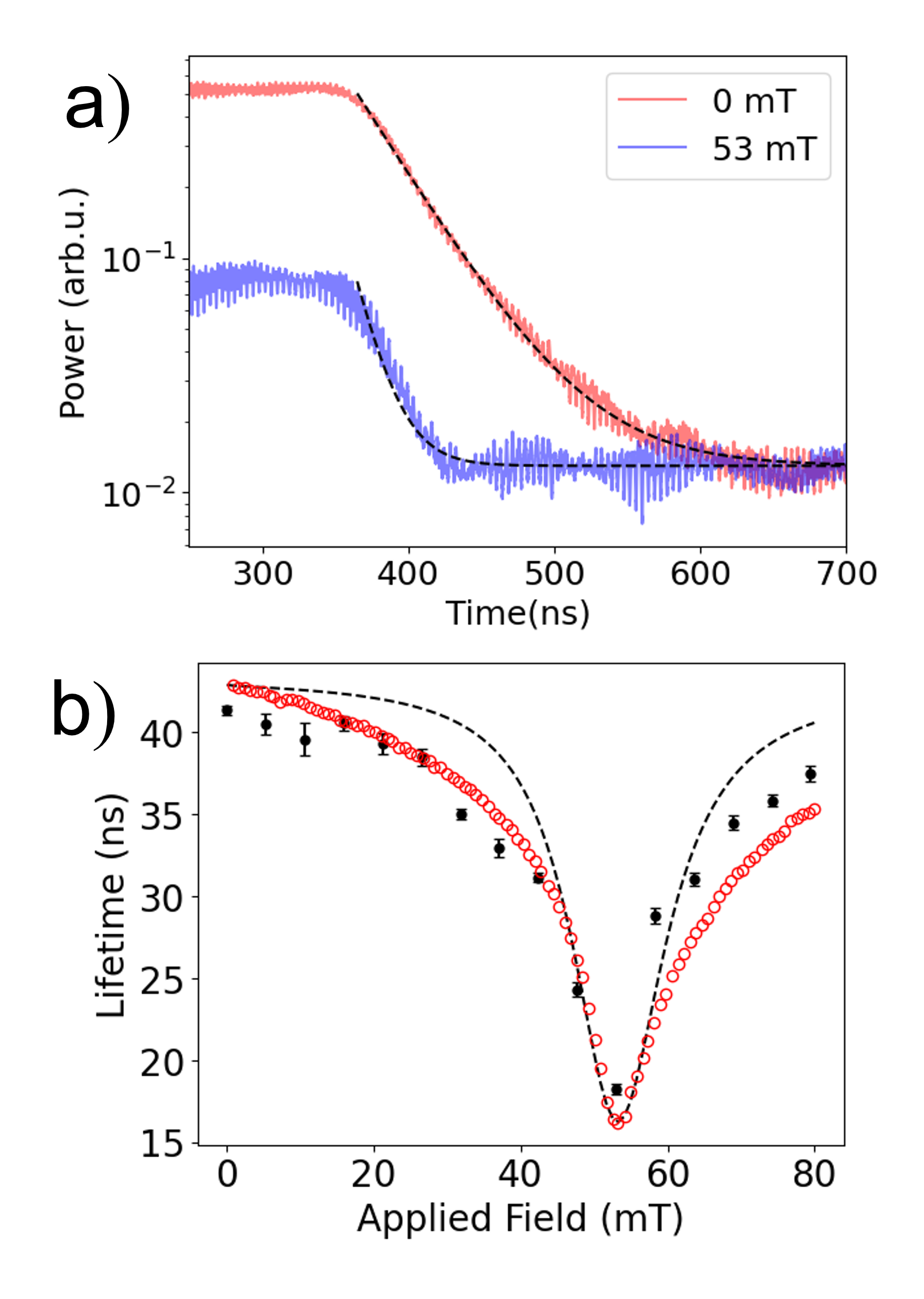}
    \caption{
    (a) Averaged transmitted power at the trailing edge of the microwave pulse for two magnetic field values: off-resonance and at resonance. Decay begins at \(t \approx \SI{362}{\nano\second}\). Dashed lines: theoretical predictions; 
    (b) Extracted photon lifetime as a function of the applied static magnetic field (black symbols). Red line: \(1/\kappa_{sys}\) obtained from frequency-domain spectroscopy. Dashed black lines: photon lifetime predicted by the Purcell model using the extracted parameters \(g\), \( \kappa_m\) and \( \kappa_c\).  
    }
    \label{fig:Cavity-ringdown}
\end{figure}

The behaviour observed in Figure~\ref{fig:Low-T-meas} shows that the system remains in the Purcell regime (Eq.~\ref{eq:RegimePurcell} even at cryogenic temperatures. We then obtain direct time-domain observation of the effect via cavity ringdown measurements~\cite{Zhang_2014}. Figure~\ref{fig:Cavity-ringdown}(a) shows 
the average transmitted power following the microwave pulse for two static magnetic fields: zero field and \SI{53}{\milli\tesla} (resonance).
A significantly faster decay is observed at resonance, indicating increased energy loss in the system.

To quantitatively assess the photon decay, we simulate the time evolution of the intra-cavity photon number \(\langle a^\dagger a \rangle\) using Eqs.~\eqref{eq:evolution} and~\eqref{eq:LindbladC} and the parameters extracted from frequency-domain measurements (Appendix~\ref{Appendix:Parameter_Cal}). The resulting curves, shown as dashed black lines in Fig.~\ref{fig:Cavity-ringdown}(a), agree quite well with the experimental data after appropriate scaling.

Since the observed decay is predominantly exponential, the photon lifetime \(\tau\) can be directly extracted from fits of the time-domain traces. Figure~\ref{fig:Cavity-ringdown}(b) shows the extracted lifetime as a function of the magnetic field. A pronounced dip is observed at resonance field \(B=\SI{53}{\milli\tesla}\), providing a clear and direct signature of the Purcell effect: the high-quality cavity's ability to store electromagnetic energy is diminished due to resonant coupling with a highly lossy magnetic excitation. The increase of decay rate \( F_P=\tau(0)/\tau(53\text{mT}) = 2.4\) , gives the Purcell factor, which from~\eqref{eq:broadening} also equals \(1+ C\), where \(C = 4g^2/\kappa_m\kappa_c\) is the cooperativity. In table \ref{tab:table_data} we display the cooperativities obtained in the frequency domain. 

Figure~\ref{fig:Cavity-ringdown}(b) also shows theoretical lifetimes determined with the aid of Eqs.~\eqref{eq:evolution} and~\eqref{eq:LindbladC} and extracted parameters from the frequency-domain spectroscopy, along with the measured values of \(1/\kappa_\mathrm{sys}\). The good agreement confirms the internal consistency of the extracted parameters across time and frequency domains.

\begin{table*}[htbp]
\caption{\label{tab:table_data}
Extracted parameters for each experimental configuration. The calculated cooperativity it is also included. Measurements (1)–(3) correspond to the same wire (w1) at different lateral positions. Measurement (7)-(8) also correspond to the same sample (w2) (wire mounted on silicon dies) in different cavities and temperatures.
}
\begin{ruledtabular}
\begin{tabular}{lccccc}
Measurement Configuration & \( \omega_c \) (GHz) & \( g/2\pi \) (MHz) & \( \kappa_m/2\pi \) (MHz) & \( \kappa_c/2\pi \) (MHz) & \( C \) \\
\hline
(1) w1 (L=4 mm)        & 7.401 & 37 & 660 & 5.6 & 1.5 \\
(2) w1 5 mm shifted         & 7.410 & 31 & 670 & 4.2 & 1.4 \\
(3) w1 8 mm shifted         & 7.425 & 19 & 680 & 2.9 & 0.7 \\
(4) L=3 mm      & 7.415 & 21 & 680 & 6.1 & 0.42 \\
(5) L=4 mm      & 7.403 & 35 & 660 & 7.6 & 0.98 \\
(6) L=5 mm      & 7.392 & 56 & 730 & 9.6 & 1.8 \\
(7) w2 (L=2.5 mm, on Si, Al cav., 300 K)         & 7.206 & 17 & 660 & 2.9 & 0.60 \\
(8) w2 (L=2.5 mm, on Si, Cu cav., 300 K)         & 7.178 & 25 & 660 & 4.1 & 0.92 \\
(9) w2 (L=2.5 mm, on Si, Cu cav., 7 mK)& 7.210 & 32 & 680 & 3.7 & 1.6 \\
\end{tabular}
\end{ruledtabular}
\end{table*}

\section{Discussions} \label{sec:Discussion}

We now turn to the consequences of wire positioning within the cavity and the enhancement of the coupling observed when the wire is placed at the electric field antinode.
Notably, for a single microwire positioned at the magnetic field antinode, the conventional configuration in cavity magnonics, no discernible shift in the cavity mode was observed above background in the VNA.
Detecting the FMR resonance in that configuration required the use of more sensitive techniques, such as phase-lock loop (PLL-FMR) setups \cite{andrade_study_2025}.
This behaviour is consistent with expectations for magnetic dipolar coupling: at the magnetic antinode, the interaction scales as 
\[
g/2\pi = \gamma B_{\mathrm{ZPF}} \sqrt{N} = \gamma \sqrt{\mu_0 \hbar \omega_c N / V},
\]
where \(V\) is the cavity mode volume.
For a  wire of length \(4\,\mathrm{mm}\), this yields an estimated coupling strength of approximately \(g/2\pi=\SI{3}{\mega\hertz}\). In contrast, the electric-field antinode configuration yields coupling rates as high as \(g/2\pi \approx \SI{37}{\mega\hertz}\), more than an order-of-magnitude enhancement.
This enhancement was further supported by measurements using larger samples containing 67 microwires positioned at the magnetic field antinode \cite{andrade_study_2025}. Comparing the induced cavity changes normalized by the number of wires (\textit{i.e.}, the number of spins), we find a 100-fold increase in the cavity amplitude response when a single wire was placed at the electric field antinode. The high quality of the cavity and the comparatively high coupling give a cooperativity \(C = 4g^2/\kappa_m \kappa_c \sim 1\) (Table \ref{tab:table_data}) for our system, validating the effectiveness of the electric-field–driven configuration.

When we measure the dependence of \( g/2\pi \) on the wire length (Fig.~\ref{fig:g-variation}), we observe a significant increase in the coupling strength with the wire length.
This enhancement can be partially understood through a simplified point-charge model proposed by Rodbell~\cite{Rodbell_ConductingPertubation}.
In this model, the induced current scales as \(\dot q \propto \dot E L^{2}\), where \(L\) is the wire length and \(\dot{E}\) is the time derivative of the electric field. This result is a circumferential magnetic field of the order 
\[
h \;\simeq\; \frac{\dot E\,L^{2}}{2\pi R},
\]
where \(R\) is the wire radius. Following this reasoning, the coupling rate can be estimated as
\[
g \;\sim\; \frac{\gamma\mu_{0}\,\omega_{c}\,E_{W,\mathrm{ZPF}}\,L^{2}\sqrt{N}}{2\pi R},
\]
where \(E_{W,\mathrm{ZPF}}\) is the zero-point electric field within the wire and \(N \propto L\) gives the number of contributing spins. This scaling implies that the absorbed power, proportional to \(\kappa_{\mathrm{sys}}\sim g^{2}\), should scale with \(L^{5}\), a trend Rodbell indeed observed~\cite{Rodbell_ConductingPertubation}. Our data appear to follow a different length dependence; only three wire lengths were measured, and more systematic studies are required to confirm the scaling.

Regarding the extracted parameters \(g\) and \(\kappa_m\) (see Figs.~\ref{fig:4mm} and \ref{fig:Low-T-meas}), we recognize that the coupled-harmonic oscillator model~(\ref{eq:Hamiltonian}) does not account for the complex nature of ferromagnetic resonance in metallic microwires, where both the susceptibility and the surface impedance are strongly influenced by the wire’s geometry and conductivity~\cite{Kraus1982,Kraus2011}. Despite these limitations, the quantitative agreement between the experimental lineshapes and the numerical simulations based on the extracted parameters, as well as the consistency between time-domain and frequency-domain measurements [Fig.~\ref{fig:Cavity-ringdown}b)], suggests that the model effectively captures the main features of the electric-field–driven FMR response in this system.

The high dissipation rate of the magnetic mode \(\kappa_m\) observed in the wires may originate from several mechanisms, including intrinsic electronic damping, anisotropy, and field-dependent eddy current losses. Remarkably, the dissipation rate remains essentially unchanged between room temperature and cryogenic conditions. Although this may seem unexpected, it could indicate that the domain damping mechanism is of electronic origin. Indeed, it has been shown~\cite{LLGultrafast} that the Gilbert parameter in metallic ferromagnets tends to saturate  below a fraction of the Curie temperature. This temperature-insensitive behaviour is consistent with our observations and suggests that this mechanism may play a key role in setting the FMR linewidth in these metallic microwires. We did not investigate the mechanisms of damping and leave it for future work. 

\section{Conclusions}

In summary, our study demonstrated that electrically mediated coupling between a magnetic microwire and a 3D microwave cavity enables observation of the Purcell effect across both room-temperature and cryogenic regimes. By placing the CoFeSiB microwire at the electric field antinode of the cavity, we accessed a regime in which microwave-induced currents generate strong localized magnetic fields that efficiently drive FMR.

Frequency-domain spectroscopy revealed cavity frequency shifts and linewidth broadening, consistent with theoretical predictions for the Purcell regime. Cavity ringdown measurements at \(T=\SI{7}{\milli\kelvin}\) provided a direct observation of the magnetically enhanced photon decay. The coupling strength exhibited a clear dependence on the position of the wire and length, supporting the interpretation of antenna-like field enhancement.

Taken together, these findings highlight both the utility and the limitations of the standard coupled-mode approach. While the Purcell regime offers a compact and effective description of the system dynamics, a more complete theory must incorporate the complex electrodynamic response of the wire, including non-uniform driving fields and current-induced effects. Future theoretical work should aim to include these contributions explicitly, potentially through spatially resolved modelling of the wire–cavity interaction.

Our results validate the use of metallic magnetic wires in cavity–magnon hybrid systems and suggest a viable route for integrating compact, high-loss materials into microwave sensing and hybrid quantum platforms. Future work could explore optimizing the wire geometry and material composition to reduce dissipation and enhance coherence, potentially enabling access to the strong-coupling regime in microscale samples.

\section{Acknowledgments}

The authors thank Dr. Alfredo Vaz for performing the FIB and SEM microscopy, and Dr. Pablo Rafael Trajano Ribeiro for insightful discussions, valuable advice, and technical support. The authors thank Prof. M. Vazquez from ICMM/CSIC Madrid for providing the wires. They also acknowledge the support provided by CCSNano/Unicamp and LAMULT/IFGW–Unicamp. Ítalo gratefully acknowledges CAPES for fellowship support (Finance Code 001-process 88887.822401/2023-00). The authors are also grateful to ESSS – Engineering Simulation and Scientific Software Ltd. for providing access to the Ansys-HFSS license, which was essential for the electromagnetic simulations carried out in this work.

This work was supported by the São Paulo Research Foundation (FAPESP) under grants 2017/08602-0, 2017/22037-4, 2017/22035-1 and 2022/07699-9; by the National Institute of Science and Technology for Micro and Nano Systems (INCT–NAMITEC, Grant 406193/2022-3); and by the Brazilian Software Excellence Promotion Association, SOFTEX (Grant 01245.002636/2023-81).

\appendix
\numberwithin{equation}{section}
\numberwithin{figure}{section}
\numberwithin{table}{section}
\section{Purcell Regime} \label{appendix:purcell_regime}

In the regime given by Equation~\ref{eq:RegimePurcell}, the density of states (DOS) of the magnetic system is very broad and can be expressed as
\begin{equation}
\text{DOS}(E) = \frac{1}{\pi\hbar}\frac{\kappa_m/2}{(\omega_m - E/\hbar)^2 + (\kappa_m/2)^2},
\end{equation}
In this limit, the transition rate of energy from the "cavity state" to the "magnet state" is described by Fermi's Golden rule~\cite{Zare_Rameshti_2022}. Since the matrix element for a single quantum transition is \(g\), the resulting cavity decay rate is given by Equation~\ref{eq:broadening}.

Equations~\ref{eq:shift} and~\ref{eq:broadening} can also be obtained directly from a model of two coupled harmonic oscillators with dissipation. In the case of classical oscillators, the system of linear equations can be written as~\cite{Zare_Rameshti_2022,Harder_cmp_lineshape}:
\begin{equation}
    \begin{cases}
    \ddot{x}_m + \kappa_m \dot{x}_m+ \omega_m^2 x_m - 2g\omega_m x_c = 0 \\
    \ddot{x}_c + \kappa_c \dot{x}_c+ \omega_c^2 x_c - 2g\omega_c x_m = 0  ,
    \end{cases}
\end{equation}
which leads to the secular equation
\begin{equation}
    \begin{vmatrix}
        \widetilde{\omega}^2 - \omega_m^2 + i \widetilde{\omega} \kappa_m & 2g\omega_m \\
        2g\omega_c & \widetilde{\omega}^2 - \omega_c^2+i\widetilde{\omega} \kappa_c
    \end{vmatrix} = 0,
\end{equation}
which can be reduced, outside the ultra-strong coupling regime ($g<<\omega_c,\omega_m$) where $\widetilde{\omega} \sim \omega_c, \omega_m$ not far from resonance, to
\begin{equation}
    \begin{vmatrix}
        \widetilde{\omega} - (\omega_m - i \kappa_m/2) & g \\
        g & \widetilde{\omega} - (\omega_c-i \kappa_c/2)
    \end{vmatrix} = 0.
\end{equation}

The complex eigenfrequencies and linewidths are then calculated to be~\cite{Topology_coupling}:
\begin{align}
    \omega_{1,2} - i \Gamma_{1,2} = \frac{1}{2} \Bigg[ 
        & \omega_c + \omega_m - i (\Gamma_c + \Gamma_m) \nonumber \\
        & \pm \sqrt{4g^2 + \left(\omega_c - \omega_m - i (\Gamma_c - \Gamma_m)\right)^2} 
    \Bigg],
    \label{eq:complete_eigenfrequencies}
\end{align}
where \( \Gamma_c=\kappa_c/2\) and \(\Gamma_m=\kappa_m/2 \) are the half-linewidths. 

The quadratic term \([\omega_c-\omega_m - i (\Gamma_c - \Gamma_m)]^2\) accounts for  resonance shifts induced by dissipation. In the strong-coupling limit, where these dissipations are negligible:
\begin{equation}
\omega_{1,2} = \frac{1}{2} \left(\omega_c + \omega_m \pm \sqrt{4g^2 + \Delta^2} \right),
\end{equation}
which are the polariton frequencies derived from the Hamiltonian~\ref{eq:Hamiltonian}.

In the general case, the real and imaginary parts of Equation~\ref{eq:complete_eigenfrequencies} are
\begin{align}
\omega_{1,2} = \frac{1}{2} \left[ (\omega_c + \omega_m) \pm \sqrt{\frac{|z|+\text{Re}(z)}{2}} \right], \label{eq:eigenfreqdissp} \\
\Gamma_{1,2} = \frac{1}{2} \left[ (\Gamma_c+\Gamma_m) \mp \text{sgn} (\text{Im}(z))  \sqrt{\frac{|z|-\text{Re}(z)}{2}} \right],
\end{align}
where \(z \equiv 4g^2 + [\omega_c-\omega_m - i (\Gamma_c - \Gamma_m)]^2 \), \(\text{Re}(z) \) and \(\text{Im}(z) \) denote respectively the real part and imaginary parts of \(z\) and \(\text{sgn} \) is the sign function. Assuming $\Gamma_m >> \Gamma_c$, one can expand $|z| (g)$ and subsequently $\sqrt{\frac{|z|\pm\text{Re}(z)}{2}}(g)$ around $g=0$ to second-order, yielding the eigenfrequencies in the Purcell regime:
\begin{align}
    \omega_{1,2} = \frac{1}{2} (\omega_c + \omega_m) \pm |\Delta| \left( \frac{1}{2}+ \frac{1}{\Delta^2 + \Gamma_m^2} g^2\right), \\
    \Gamma_{1,2} = \frac{1}{2} (\Gamma_c + \Gamma_m) \pm \text{sgn}(\Delta) \left( \frac{(\Gamma_m-\Gamma_c)}{2}- \frac{\Gamma_m}{\Delta^2 + \Gamma_m^2} g^2\right).
\end{align}

\begin{figure}[h]
    \centering
    \includegraphics[width=0.8\linewidth]{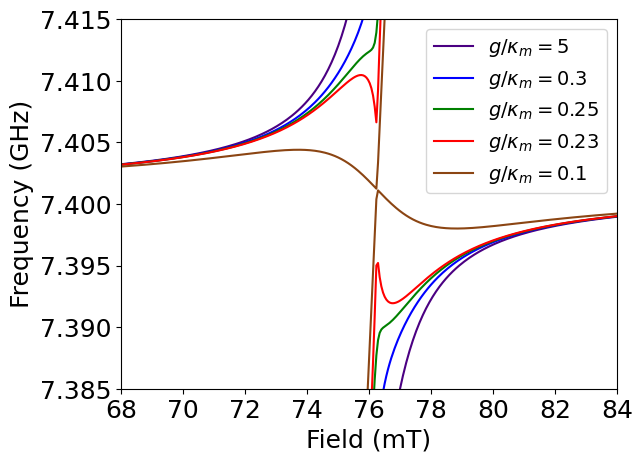}
    \caption{
    Eigenfrequencies calculated from Equation~\ref{eq:complete_eigenfrequencies} for different values of $\kappa_m$. The parameters used were: $g/2\pi=5  \,\text{MHz}$, $\kappa_c/2\pi=5  \,\text{MHz}$, $\omega_c/2\pi = 7.4 \, \text{GHz}$ and $\omega_m = \gamma \sqrt{B(B+0.84\text{T})}$.
    }
    \label{fig:transition_to_Purcell}
\end{figure}

As $\Delta$ crosses zero, each branch rapidly changes to a pure magnetic character, so that the observable cavity response switches branches. This behaviour is captured by Equations~\ref{eq:shift} and~\ref{eq:broadening}. Figure~\ref{fig:transition_to_Purcell} shows the eigenfrequencies (Equation~\ref{eq:eigenfreqdissp}) as a function of the detuning for several values of \(\kappa_m\), illustrating the transition to the Purcell regime. We also note that \(|z|\) and \(-\text{Re}(z)\) correspond respectively to the quantities \(\delta'\) and \(\delta\) defined in~\cite{zhao2025theorymagnonpurcelleffect}, which appear in the  eigenvalues of the matrix \(\hat{\Omega}\).
 
\section{Calculation of Parameters and Decay}\label{Appendix:Parameter_Cal}

The parameters \( \omega_c \), \( \kappa_c \), \( \kappa_m \), and \( g \) were extracted from the experimental data using Equations~\eqref{eq:shift} and~\eqref{eq:broadening}.
In the frequency shift curve (\textit{e.g.} Figure~\ref{fig:4mm}(c)), the separation between the extrema (the “waist”) corresponds to \( \kappa_m \), while the difference between the maximum and minimum values (the “height”) equals \( 2g^2/\kappa_m \).
The asymptotic value of \( \omega_{\mathrm{sys}} \;\simeq\; \omega_c - g^2/\Delta \) for large detuning (\( \Delta \gg \kappa_m \)) provides the bare cavity frequency \( \omega_c \).
Similarly, the absorption curves yield \( \kappa_c \), \( g \) and \( \kappa_m \): the baseline corresponds to \( \kappa_c \), the peak height to \( 4g^2/\kappa_m \), and the full width at half maximum (FWHM) to \( \kappa_m \). 
These quantities were extracted from interpolated data when needed. 
Since the coupled-oscillator model assumes symmetric Lorentzian lineshapes, which do not fully capture the experimentally observed asymmetries, the extracted parameters represent effective values within the harmonic approximation.

Alternative parameter fitting was previously performed in~\cite{andrade_study_2025}, using a least-squares fit near resonance. That procedure yielded comparable values for \( g/2\pi \), but higher magnetic dissipation (\( \kappa_m/2\pi \sim \SI{800}{\mega\hertz} \)).
The fitting emphasized agreement with the Lorentzian tails, which are overrepresented in the data due to sampling density. Because the measured lineshapes decay more slowly than a Lorentzian, these fits tend to overestimate \( \kappa_m \) and may require artificial values of \( \kappa_c \) unless the full curve is fitted which gives even higher values of \(\kappa_m\). For this reason, the height-waist method used in this study is preferred, offering a simpler and more physically consistent approach. 

For the decay curves shown in Figure~\ref{fig:Cavity-ringdown}(a), the photon number dynamics were computed from Equations~\eqref{eq:evolution} and~\eqref{eq:LindbladC} using the matrix exponential \( e^{\Omega t} \), with parameters obtained from frequency-domain fits. Initial conditions assumed no initial magnon population (\( \langle m^\dagger m \rangle = 0 \)), no coherence between magnon and photon (\( \langle m^\dagger a \rangle = 0 \)), and an initial cavity population \( \langle a^\dagger a \rangle = N_a \). In the Purcell regime, the resulting evolution follows approximately \( \langle a^\dagger a \rangle \propto e^{-\kappa_{\mathrm{sys}} t} \).

To match the numerical decay curves with experimental data, we rescaled the evolution using the average transmitted power \( A \) during the flat part of the pulse plateau (\( t = 0 \rightarrow 250~\text{ns} \)), and added a background offset equal to the mean power at the noise floor (\( t = 600 \rightarrow 1000~\text{ns} \)). Thus, the simulated curves shown in Figure~\ref{fig:Cavity-ringdown}(a) correspond to:
\[
P_{\text{theory}}(t) = A \, [e^{\Omega t}]_{11} + 0.013.
\]
The theoretical photon lifetimes in Figure~\ref{fig:Cavity-ringdown}(b) were obtained by fitting single exponential decays to the simulated time traces.

\section{Cryogenic Measurement System}\label{appendix:cryogenic_measurement}

\begin{figure}[ht]
    \centering
    \includegraphics[width=0.8\linewidth]{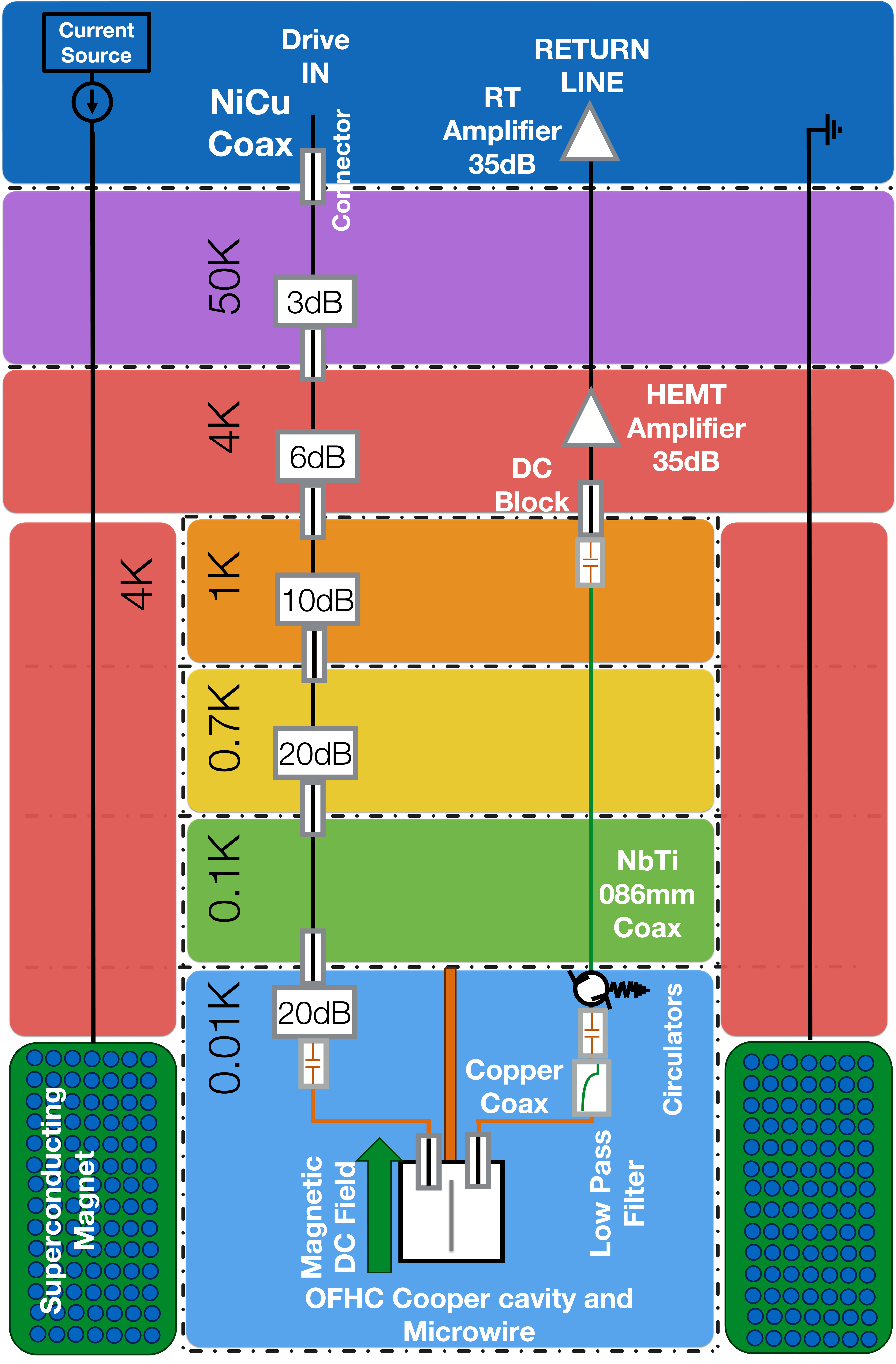}
    \caption{
    Schematic of the cryogenic measurement setup implemented in the \textit{BlueFors} LD dilution refrigerator. The copper cavity is thermally anchored to the MXC and positioned within the bore of a superconducting magnet mounted on the \SI{4}{\kelvin} stage. Coaxial lines with distributed cryogenic attenuation and isolation stages connect room-temperature instruments to the sample. Signal amplification is provided by a cryogenic HEMT amplifier and a secondary low-noise amplifier at room temperature.
    }
    \label{fig:sup-cryostat}
\end{figure}

Cryogenic measurements were performed using a commercial \textit{BlueFors} LD dry dilution refrigerator with a base temperature of approximately \SI{7}{\milli\kelvin} (see Fig.~\ref{fig:sup-cryostat}). 
A superconducting magnet (American Magnetics), thermally anchored to the \SI{4}{\kelvin} stage, was used to apply a static magnetic field along the vertical axis of the refrigerator. 
Current was supplied to the magnet through superconducting leads spanning the \SI{50}{\kelvin} and \SI{4}{\kelvin} stages, in order to minimize ohmic dissipation and thermal loading.
The magnet supports fields up to \SI{8}{\tesla}, with a control resolution better than \SI{5d-4}{\tesla}, and it is driven by a quadrupolar current source operated through a programmable interface on a local network. Magnet control was automated using a custom Python-based script interfaced via the \texttt{socket} library.

To ensure alignment between the sample and the magnetic field centre, located approximately \SI{400}{\milli\meter} below the MXC  flange, a dedicated sample holder was developed. Since the aluminium cavity used at room temperature becomes superconducting at cryogenic temperatures, screening the external magnetic field, a new cavity was fabricated from OFHC copper. This cavity maintained the same internal geometry as the aluminium version but housed a larger enclosure to allow vertical mounting via screw holes along the \( y \)-axis. A custom copper bracket attached the cavity to a rigid copper rod aligned with the magnetic axis, ensuring spatial overlap between the cavity centre and the magnetic centre. The copper rod was mechanically  and thermally anchored to the MXC flange using a machined clamp to ensure good thermalization and mechanical stability.

Microwave lines were attenuated at multiple temperature stages, as shown in Fig.~\ref{fig:sup-cryostat}, resulting in an estimated input power of \SI{1.3d-10}{\watt} at the cavity. On the output path, two cryogenic circulators provided approximately \SI{30}{\decibel} of isolation between the cavity and the first-stage amplifier: a cryogenic high-electron-mobility transistor (HEMT) amplifier with \SI{35}{\decibel} gain mounted on the \SI{4}{\kelvin} stage, followed by a second room-temperature low-noise amplifier with an additional \SI{35}{\decibel} gain.

\section{Cavity Ringdown Measurement}\label{Appendix:ringdown}

Cavity ringdown measurements were performed at cryogenic temperatures using a home-built superheterodyne detection system, adapted from existing microwave instrumentation developed for superconducting qubit experiments. The setup allows direct time-domain observation of signals transmitted through the cavity.
A schematic of the measurement chain is shown in Figure~\ref{fig:sup-RT_and_FAST_Aquisition}(c). An arbitrary waveform generator (AWG) was used to modulate a continuous-wave microwave generator operating at \SI{6.97}{\giga\hertz}, producing square pulses of \SI{200}{\micro\second} duration. The modulated signal was mixed with a \SI{240}{\mega\hertz} local oscillator to generate the excitation tone at the cavity frequency (\(\sim\SI{7.21}{GHz}\)), which was then sent to the dilution refrigerator. Although the system involved up-conversion, for the purposes of these measurements, it effectively acted as a pulsed source near \SI{7.1}{\giga\hertz}. The transmitted signal was amplified at cryogenic and room temperature, down-converted using a second mixer and local oscillator to the same \SI{240}{\mega\hertz} intermediate frequency, and digitized by a high-speed (20 GSa/s) oscilloscope triggered by the AWG. For each applied magnetic field value, 100 transmission traces were acquired and square-averaged to obtain the time-resolved average transmitted power of the microwave pulses.

\bibliographystyle{apsrev4-2}

\bibliography{biblio.bib}

\end{document}